\documentclass{article}
\usepackage{graphicx}
\usepackage{html}
\begin{document} 
\bibliographystyle{prsty}
\title{A Mechanism for Spatio-Temporal Disorder in Bistable Reaction-Diffusion
Systems} 

\author{
\htmladdnormallink{Aric Hagberg}{http://t7.lanl.gov/People/Aric}\\
\htmladdnormallink{Center for Nonlinear Studies}{http://cnls-www.lanl.gov} 
                   and 
\htmladdnormallink{T-7,}{http://t7.lanl.gov} 
\htmladdnormallink{Theoretical Division,}{http://www-tdo.lanl.gov}\\
\htmladdnormallink{Los Alamos National Laboratory,}{http://www.lanl.gov}
                   Los Alamos, NM 87545\\
\htmladdnormallink{(aric@lanl.gov)}{mailto:aric@lanl.gov}\\\\\latex{\and}
\htmladdnormallink{Ehud Meron}{http://www.bgu.ac.il/BIDR/research/staff/meron.html}\\
\htmladdnormallink{The Jacob Blaustein Institute for Desert Research}{http://www.bgu.ac.il/BIDR}		
                   and the 
\htmladdnormallink{Physics Department}{http://www.bgu.ac.il/phys/physics.html},\\
\htmladdnormallink{Ben-Gurion University}{http://www.bgu.ac.il}, 
		   Sede Boker Campus 84990, Israel\\
\htmladdnormallink{(ehud@bgumail.bgu.ac.il)}{mailto:ehud@bgumail.bgu.ac.il}\\
}
\date{\today}
\maketitle

\html{$\mbox{For best results, please set your browser to the width and font size of this line of text}$}

\latex{The electronic form of this document can be found at http://www.springer-ny.com/nst/nstarticles.html}

\begin{abstract}
In bistable systems, the stability of front structures 
often influences the dynamics of extended patterns.  We show
how the combined effect of an instability to curvature
modulations and proximity to a pitchfork front bifurcation leads to spontaneous
nucleation of spiral waves along the front. This effect is demonstrated by
direct simulations of a FitzHugh-Nagumo (FHN) model and by simulations of
order parameter equations for the front velocity and curvature. 
Spontaneous spiral-wave nucleation
often results in a state of spatio-temporal disorder involving
repeated events of spiral wave nucleation, domain spliting and spiral wave
annihilation. 
\end{abstract}

\newpage

\section{Introduction}
Spatio-temporal disorder is a ubiquitous phenomenon in extended nonequilibrium
systems yet the mechanisms responsible for it are only partially explored.
A generic mechanism for the onset of disorder in {\em periodic patterns} 
consists of phase instabilities followed by the 
formation of phase singularities which appear as
dislocation defects or vortices \cite{CGLe:89b}. This mechanism has been 
observed in numerical simulations of the complex Ginzburg-Landau equation 
\cite{CGLe:89,Bohr:90}, and in 
experiments, e.g. electroconvection in liquid crystals \cite{RRS:89}. 
A considerable effort has been devoted to elucidating the nature of the
transition from the regular phase regime to the regime where vortices or
defects spontaneously appear~\cite{SPS:92,GrEg:95,GR:96,MC:96,CLMM:96}. 
Other mechanisms involving
instabilities of periodic patterns, leading to spiral breakup, have been
reported recently \cite{CoWi:91,HoPa:91,Karm:93,BE:93}.

In this article we present a mechanism for the onset of
spatio-temporal disorder associated with {\em front structures}.
This type of disorder is illustrated by a
numerical solution of a bistable FitzHugh Nagumo (FHN) type 
reaction-diffusion system
(\hyperref{Figure 1}{Fig. }{}{figure:front_turb}).
An almost flat front, connecting the two stable
uniform stationary states, begins traveling through the
system.  The front represents an ``up'' state
\latex{(black)}\html{(red)} invading a ``down''
state (white).  Initial nonuniform perturbations of the front
position grow into wiggles which nucleate pairs of spiral waves 
(see the 
\htmlref{Appendix}{appendix}
for a different view of the middle frame in 
\hyperref{Figure 1}{Fig. }{}{figure:front_turb}.)
The solution then evolves into a disordered state
with repeated events of domain splitting and spiral-wave
nucleation and annihilation. Similar phenomena have been
observed in experiments on the ferrocyanide-iodate-sulfite
(FIS) reaction \cite{LMSP:94,LeSw:95}.

\begin{latexonly}
\begin{figure}
\centering\includegraphics[width=5.5in]{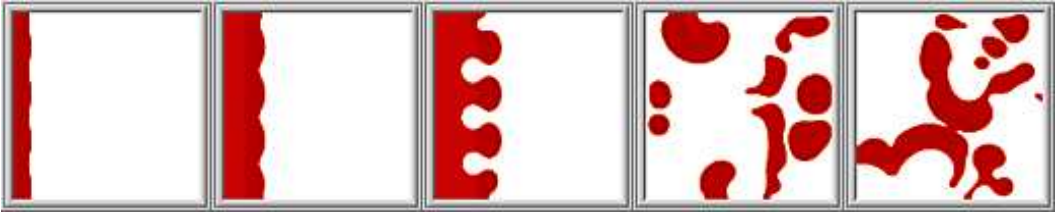}
\caption{
The evolution of 
an unstable front solution connecting the up state (black) 
to the down state (white) in a bistable FHN reaction-diffusion system.
The frames, from left to right, represent the pattern solution
at successive moments in time.
Perturbations on the flat traveling front grow and nucleate
spiral pairs along the front line  
(see the 
\htmlref{Appendix}{appendix}
for a different view of the middle frame).
The resulting disordered
state is characterized by domain splitting and spiral-wave
nucleation and annihilation.
}
\label{figure:front_turb}
\end{figure}
\end{latexonly}

\begin{htmlonly}
\label{figure:front_turb}
\begin{rawhtml}
<center>
<a href="front8.gif">
<img src=front_static.gif></a>
<br>
<a href="front8.gif">[Animated GIF] </a>
<a href="front8.mpg">[MPEG] </a>
<br>
<br>
<STRONG>Figure 1:</STRONG>
\end{rawhtml}
The evolution of 
an unstable front solution connecting the up state (red) 
to the down state (white) in a bistable FHN reaction-diffusion system.
The frames, from left to right, represent the pattern solution
at successive moments in time.
Perturbations on the flat traveling front grow and nucleate
spiral pairs along the front line
(see the 
\htmlref{Appendix}{appendix}
for a different view of the middle frame).
The resulting disordered
state is characterized by domain splitting and spiral-wave
nucleation and annihilation.
\begin{rawhtml}
<BR>
</center>
\end{rawhtml}
\end{htmlonly}

\section{The NIB front bifurcation and spontaneous front reversals}

The key to understanding the disorder associated with front
structures is a pitchfork bifurcation in the velocity of
a propagating front.  The front bifurcation is represented by the equation
\begin{equation}
C^3-(\lambda_c-\lambda)C=0\,,
\label{pfork}
\end{equation}
where $C$ is the velocity of a flat front and $\lambda$ is a
control parameter.  The corresponding bifurcation diagram is
shown in \hyperref{Figure 2}{Fig. }{}{figure:frontbif}.
A stationary front ($C=0$) becomes unstable below a
critical parameter value, $\lambda=\lambda_c$. 
At that point two new stable front
solutions appear, representing an up state invading a down state ($C>0$) and a
down state invading an up state ($C<0$).  Hereafter, we refer to these front
solutions as to ``UD front'' and ``DU front'', respectively.

This type of front bifurcation has been derived for periodically forced 
oscillatory systems \cite{CLHL:90} and for bistable FHN type models 
\cite{IMN:89,HaMe:94a,BRSP:94}.  Experimental observations of front
bifurcations, or supporting evidence for their existence, have been reported 
in Refs. \cite{FRCG:94,NYK:95} for liquid crystals, and in Refs. 
\cite{HBKR:95,Haim:96} for chemical reactions.
The stationary and counter-propagating fronts
are sometimes referred to as Ising and Bloch fronts, respectively, and the
bifurcation itself as a nonequilibrium Ising-Bloch (NIB) bifurcation
\cite{CLHL:90,HaMe:94a}.

\begin{rawhtml}
<center>
\end{rawhtml}
\begin{figure}
\htmlimage{align=center,thumbnail=1.25}
\centering\includegraphics[width=4.5in]{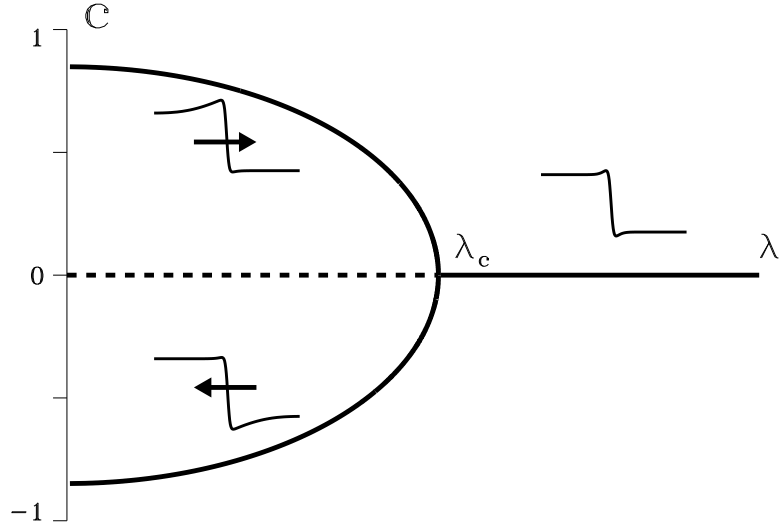}
\caption{
The pitchfork front bifurcation. 
At $\lambda=\lambda_c$ the stationary ($C=0$) front solution
becomes unstable to a pair of counterpropagating travelling fronts.
}
\label{figure:frontbif}
\end{figure}
\begin{rawhtml}
</center>
\end{rawhtml}

Physical realizations of front
bifurcations usually involve perturbations that unfold the pitchfork form 
of (\ref{pfork}) into \cite{GoSh:85}
\begin{equation}
C^3-(\lambda_c-\lambda)C+\nu_1+\nu_2 C^2=0\,.
\label{ppfork}
\end{equation}
An asymmetry between the up and down states is an example of such a
perturbation. In the following we confine ourselves to the case $\nu_2=0$. 
A plot of the 
surface (\ref{ppfork}) in the space spanned by $C,\lambda$ and $\nu_1$ 
is shown in 
\hyperref{Figure 3}{Fig. }{}{figure:c-lambda-nu}.
The significance of small variations of $\nu_1$ in the vicinity of the
pitchfork bifurcation point, $\lambda=\lambda_c$, $\nu_1=0$, is now evident:
perturbations may induce transitions between the upper and lower 
sheets and reverse the direction of front propagation.
Notice that farther from the bifurcation 
point the variations of  $\nu_1$ must be larger in order to induce 
front reversal.

\begin{latexonly}
\begin{figure}
\centering\includegraphics[width=4.5in]{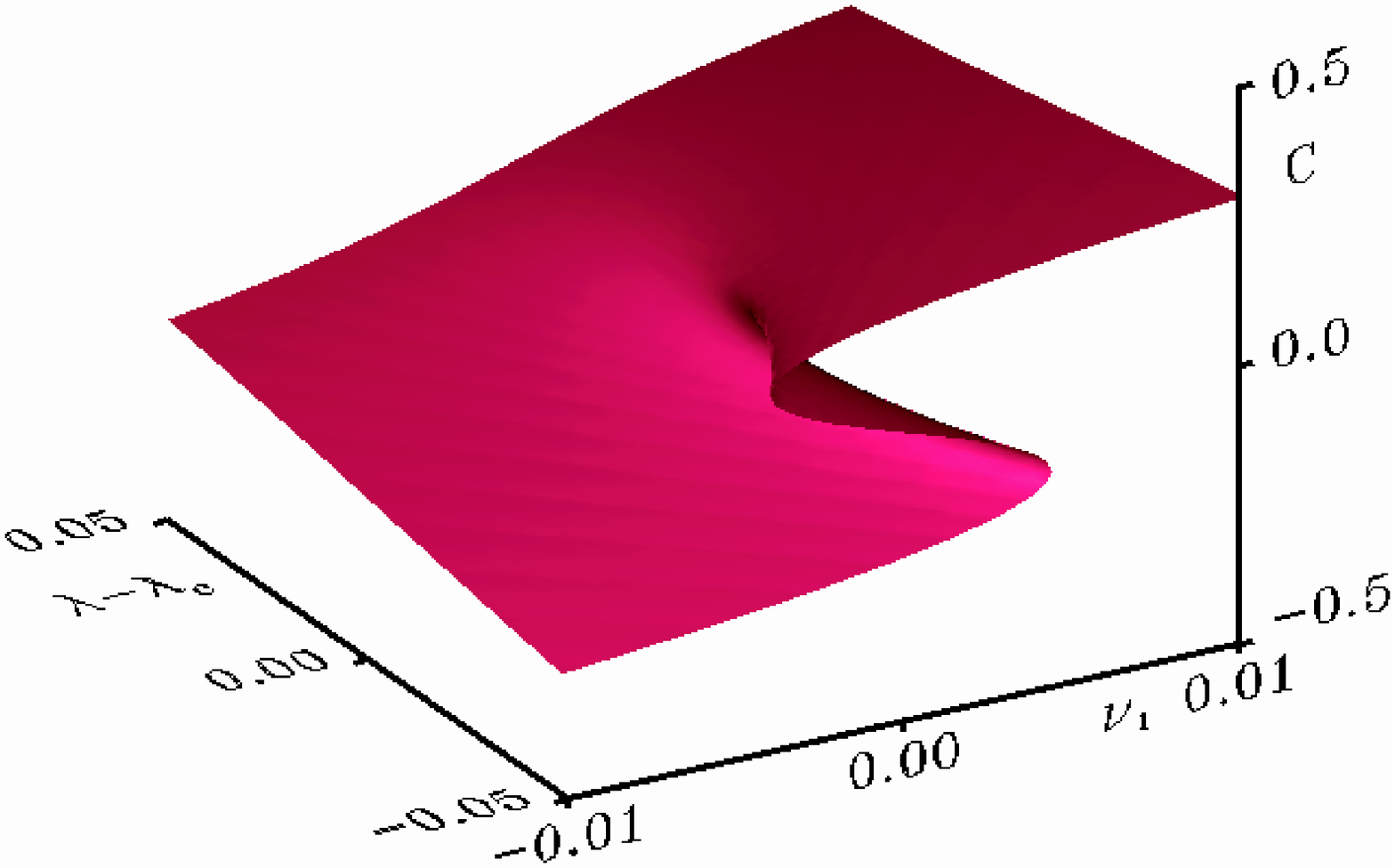}
\centering\includegraphics[width=7.0in]{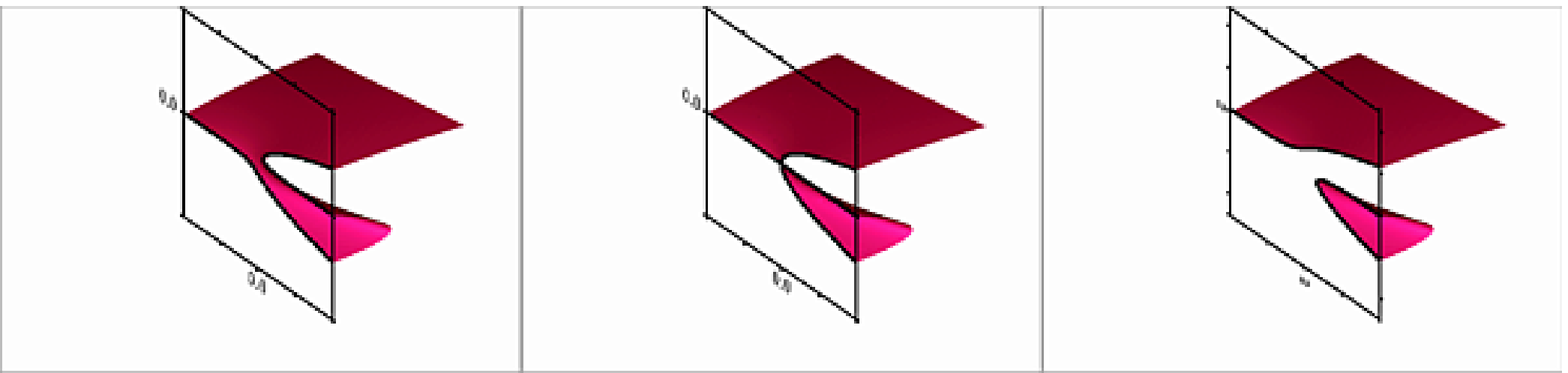}
\centering\includegraphics[width=7.0in]{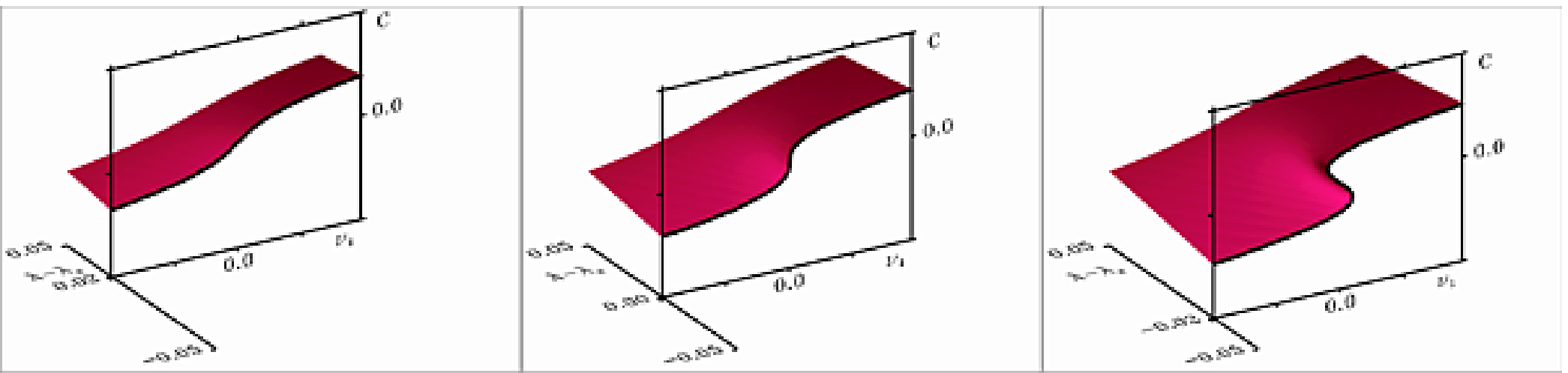}
\caption{
Top: The surface $C^3-(\lambda_c-\lambda)C+\nu_1=0$ in the space 
spanned by $C,\lambda,\nu_1$ (the cusp catastrophe).
First triad: A section at $\nu_1=0$ showing the pitchfork 
bifurcation (center), and sections at $\nu_1<0$ (left) and $\nu_1>0$ (right)
showing unfoldings of the pitchfork. Second triad: Sections of the surface at 
constant $\lambda$ showing the hysteresis 
point (center) with single valued
(left) and multivalued (right) relations
away from the hysteresis point.
}
\label{figure:c-lambda-nu}
\end{figure}
\end{latexonly}

\begin{htmlonly}
\label{figure:c-lambda-nu}
\begin{rawhtml}
<center>
<STRONG>Figure 3:</STRONG>
<br>
<a href="surface_t.gif">
<img src=surface_t_thumb.gif></a>
<br>
\end{rawhtml}
The surface $C^3-(\lambda_c-\lambda)C+\nu_1=0$ in the space 
spanned by $C,\lambda,\nu_1$ (the cusp catastrophe).
\begin{rawhtml}
<br>
<a href="saddle_down_t.gif">
<img src=saddle_down_t_thumb.gif></a>
<a href="pitchfork_t.gif">
<img src=pitchfork_t_thumb.gif></a>
<a href="saddle_up_t.gif">
<img src=saddle_up_t_thumb.gif></a>
<br>
\end{rawhtml}
A section at $\nu_1=0$ showing the pitchfork 
bifurcation (center), and sections at $\nu_1<0$ (left) and $\nu_1>0$ (right)
showing unfoldings of the pitchfork.
\begin{rawhtml}
<br>
<a href="single_t.gif">
<img src=single_t_thumb.gif></a>
<a href="hyster_t.gif">
<img src=hyster_t_thumb.gif></a>
<a href="multi_t.gif">
<img src=multi_t_thumb.gif></a>
<br>
\end{rawhtml}
Sections of the surface at 
constant $\lambda$ showing the hysteresis 
point (center) with single valued
(left) and multivalued (right) relations
away from the hysteresis point.
\begin{rawhtml}
<br>
<BR>
</center>
\end{rawhtml}
\end{htmlonly}

The dynamics of a single flat front may also be changed by slow dynamical
processes such as an approach to
a boundary, an interaction with another front, or development of
curvature. 
\hyperref{Figure 4}{Fig. }{}{figure:c-k}
shows a typical graph of the normal velocity of a front,
$C_n$, versus its curvature, $\kappa$, for fixed $\lambda$ 
near the front bifurcation. 
The figure was obtained using a singular perturbation analysis of
an FHN model, assuming slow curvature dynamics with respect to the time scale
of front reversal \cite{HaMe:94c}.
The hysteretic shape, similar to the graph of $C$ versus $\nu_1$
for fixed $\lambda$ (see \hyperref{Figure 3}{Fig. }{}{figure:c-lambda-nu}), 
suggests that front reversals can be induced by 
small curvature perturbations.
Since curvature is an intrinsic dynamical variable, reversals of 
propagation direction occur spontaneously as the curvature
of a front changes.
Spontaneous front reversals in catalytic reactions on platinum
surfaces have indeed been observed for parameters
in the vicinity a front bifurcation
\cite{HBKR:95}. Experimental observations of front reversals induced by
boundaries have been reported in \cite{Haim:96}.

\begin{rawhtml}
<center>
\end{rawhtml}
\begin{figure}
\htmlimage{thumbnail=1.25,align=center}
\centering\includegraphics[width=4.5in]{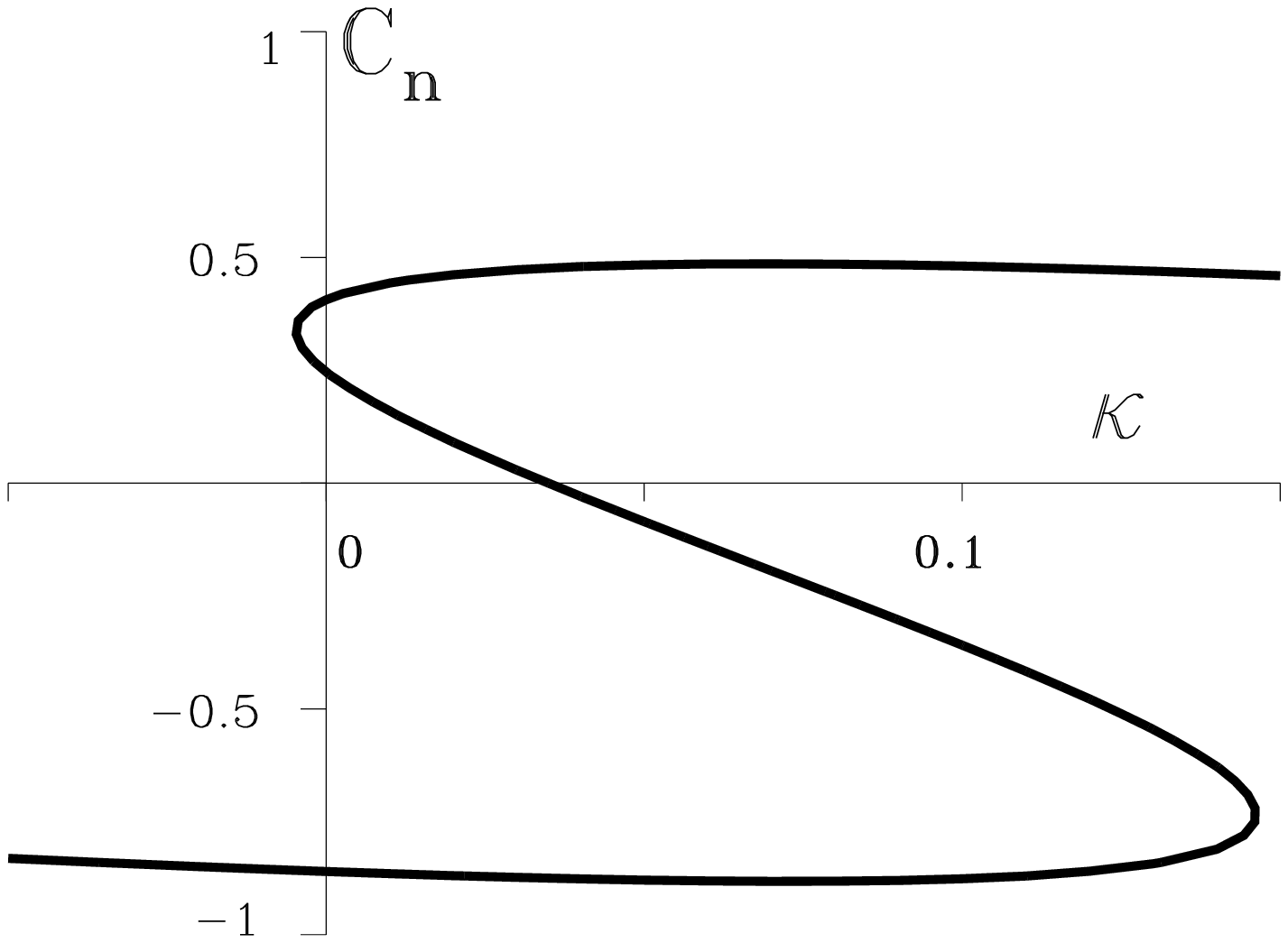}
\caption{
Normal front velocity $C_n$ $vs$  curvature $\kappa$ 
in the vicnity of the front bifurcation.
The development of small 
negative curvature may induce a transition from a UD front ($C_n>0$) to a 
DU front ($C_n<0$).             
}
\label{figure:c-k}
\end{figure}
\begin{rawhtml}
</center>
\end{rawhtml}

Imagine now a flat UD front that
is unstable to transverse perturbations (i.e. an instability to curvature
perturbations). As the front propagates, alternating segments along the front
acquire negative and positive curvatures.  For parameters that
place the system near the the front bifurcation,
where 
\hyperref{Figure 4}{Fig. }{}{figure:c-k}
applies, segments with negative curvature will 
eventually reverse
propagation direction. Such local front reversals involve the creation of
transition zones between the counterpropagating UD and DU fronts. These zones 
form the cores of {\em rotating spiral waves}.

The scenario sketched above provides a heuristic explanation for the 
spontaneous nucleation of spiral waves in 
\hyperref{Figure 1}{Fig. }{}{figure:front_turb}
which precedes the onset of spatio-temporal disorder. Similar arguments hold
for other intrinsic perturbations such as front interactions. Indeed 
\hyperref{Figure 1}{Fig. }{}{figure:front_turb}
includes many events where local front reversals are
induced by the interactions between approaching domains. An example of such an
event is shown in 
\hyperref{Figure 5}{Fig.~}{}{figure:front_inset}.
Similar processes have been observed in
\hyperref{experiments}
{experiments (Figure }  
{)}
{figure:cnld}
on the FIS reaction \cite{LMSP:94,LeSw:95}.

\begin{latexonly}
\begin{figure}
\centering\includegraphics[width=5.5in]{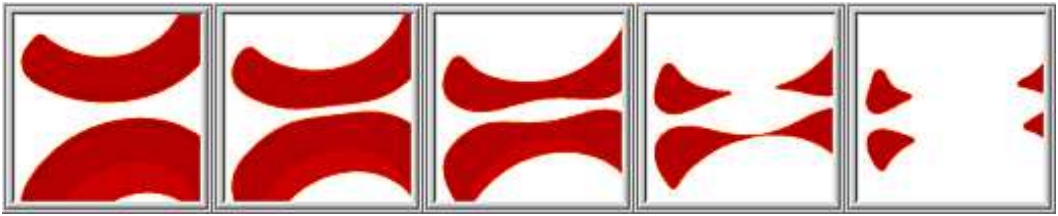}
\caption{
A closer look at front interactions in the numerical solution
of the FHN model from 
\hyperref{Figure 1}{Fig. }{}{figure:front_turb}.
The repulsive interaction between approaching fronts
causes them to reverse direction.
The reversal is followed by domain splitting.
}
\label{figure:front_inset}
\end{figure}
\end{latexonly}

\begin{htmlonly}
\label{figure:front_inset}
\begin{rawhtml}
<center>
<a href="inset.gif">
<img src=inset_static.gif></a>
<br>
<a href="inset.gif">[Animated GIF] </a>
<a href="inset.mpg">[MPEG] </a>
<br>
<br>
<STRONG>Figure 5:</STRONG>
\end{rawhtml}
A closer look at front interactions in the numerical solution
of the FHN model from 
\hyperref{Figure 1}{Fig. }{}{figure:front_turb}.
The repulsive interaction between approaching fronts
causes them to reverse direction.
The reversal is followed by domain splitting.
\begin{rawhtml}
<BR>
</center>
\end{rawhtml}
\end{htmlonly}

\begin{latexonly}
\begin{figure}
\centering\includegraphics[width=5.5in]{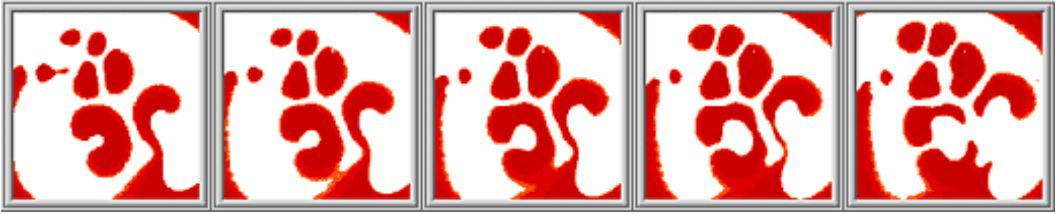}
\caption{
Patterns in the Ferrocyanide-Iodate-Sulfite reaction show 
interactions leading to front reversals
followed by domain splittings.  These
images are from experiments performed in the 
\protect\htmladdnormallink{Center for Nonlinear Dynamics}
{http://chaos.ph.utexas.edu}
at the University of Texas at Austin.
}
\label{figure:cnld}
\end{figure}
\end{latexonly}

\begin{htmlonly}
\label{figure:cnld}
\begin{rawhtml}
<center>
<a href="cnld.gif">
<img src=cnld_static.gif></a>
<br>
<a href="cnld.gif">[Animated GIF] </a>
<a href="cnld.mpg">[MPEG] </a>
<br>
<br>
<STRONG>Figure 6:</STRONG>
\end{rawhtml}
Patterns in the Ferrocyanide-Iodate-Sulfite reaction show 
interactions leading to front reversals
followed by domain splittings.  These
images are from experiments performed in the 
\htmladdnormallink{Center for Nonlinear Dynamics}
{http://chaos.ph.utexas.edu}
at the University of Texas at Austin.
\begin{rawhtml}
<BR>
</center>
\end{rawhtml}
\end{htmlonly}

\section{The dynamics of front reversals}

Algebraic relations like that displayed in  
\hyperref{Figure 4}{Fig. }{}{figure:c-k}
are useful for predicting
the onset of spiral-wave nucleation:  nucleation events 
become feasible when a single valued relation becomes multivalued (or
hysteretic). Such relations do not, however, contain information about the
nucleation process itself.  To study the nucleation process 
a time-dependent approach for the front dynamics is needed.
Near the bifurcation, the asymptotic front dynamics are governed by both a
translational degree of freedom and an 
order parameter associated with the bifurcation:
the front velocity $C$.  
In Refs. \cite{EHM:95,HMRZ:96} we derived
asymptotic dynamic equations for a single unperturbed one-dimensional
front using FHN type models.  The equations for the
front position, $X$, and velocity are
\begin{eqnarray}
\dot X &=& C \,, \\ \nonumber
\dot C &=& (\alpha_c-\alpha)C -\beta C^3 \,.
\label{normal}
\end{eqnarray}
These uncoupled equations describe the convergence to constant speed motion 
Ising (zero speed) and Bloch fronts. 
Order parameter equations for front
propagation in one space dimension near a NIB bifurcation have also been
derived in Ref.~\cite{Bode:96} to study the effects of external fixed
heterogeneities.

The effect of curvature 
is to couple the two equations in a way that allows for front
reversal. The equations for the dynamics of a single 
two-dimensional front
with smooth curvature, $\kappa$, are \cite{HaMe:96}
\begin{equation}
{\partial\kappa\over\partial t} = -(\kappa^2 + {\partial^2\over\partial
s^2})C_n - {\partial\kappa\over\partial s}\int_0^s \kappa C_n ds^\prime \,,
\label{K}
\end{equation}
and 
\begin{eqnarray}
{\partial C\over\partial t}&=&(\alpha_c-\alpha)C -\beta C^3
+\gamma\kappa + \gamma_0  
+{\partial^2 C\over \partial s^2} 
- {\partial C \over \partial s} \int_0^s \kappa C_n ds^\prime\,, 
\label{C}
\end{eqnarray}
where $C_n$, the normal front velocity, is given by
\begin{equation}
C_n = C - D\kappa\,,
\nonumber
\end{equation}
and $s$ is the arclength coordinate along the front. In deriving these
equations an asymmetry between the up and down states has been introduced.
Equation (\ref{K}), for the curvature of the front, 
follows from purely geometric considerations 
\cite{Mikhailov:90,Meron:92}.  Equation
(\ref{C}), for the speed of the front,
is valid near the front bifurcation and the boundary
of instability to transverse perturbations \cite{HaMe:94c,HaMe:96}. 
The integral term in both equations 
represents ``advection'' of changes in $C$
and $\kappa$ from the stretching of the arclength over time.
Note that away from the front bifurcation where the time scale associated with
front reversal, $(\alpha_c-\alpha)^{-1}$, is short, $C$ is no longer a slow
variable and can be eliminated adiabatically.  For a circular front, equation
(\ref{C}) then reduces to an algebraic relation between the normal front
velocity and its curvature such as the one in
\hyperref{Figure 4}{Fig. }{}{figure:c-k}.

We have computed numerical solutions of equations (\ref{K}) and (\ref{C}) 
starting with an almost flat UD front as 
an initial condition.  
\hyperref{Figure 7}{Fig. }{}{figure:ck_stable}
pertains to parameter values
in the Bloch regime where the UD and DU fronts coexist
and are both stable to transverse perturbations. The initial front
converges to a flat UD front propagating at constant speed.
\begin{latexonly}
\begin{figure}
\centering\includegraphics[width=5.5in]{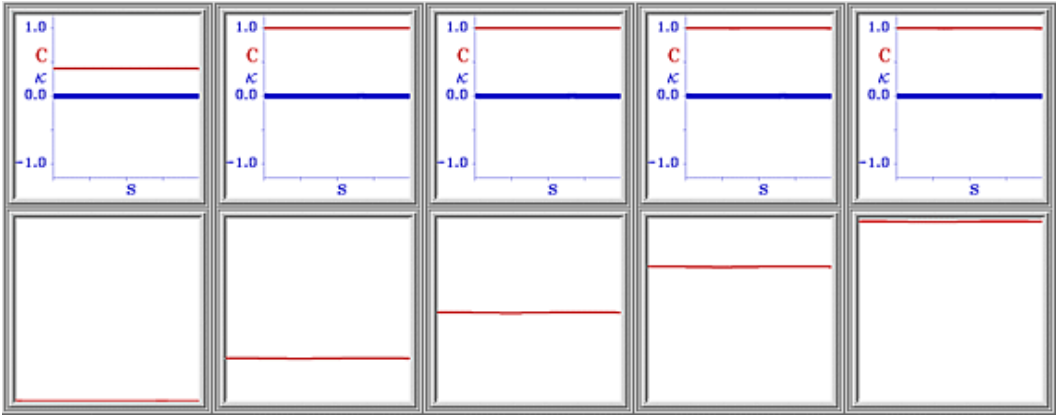}
\caption{
A solution to equations (\protect\ref{K}) and (\protect\ref{C}) 
when the two-dimensional front is stable to transverse perturbations.
The frames in the top row show the time evolution of 
$C(s)$ (thin line) and $\kappa(s)$ (thick line). 
The solution converges to a constant
speed flat ($\kappa=0$) traveling front.  The frames in the bottom
row display the corresponding dynamics of the front in 
the physical two-dimensional plane.
}
\label{figure:ck_stable}
\end{figure}
\end{latexonly}
\begin{htmlonly}
\label{figure:ck_stable}
\begin{rawhtml}
<br><br>
<center>
<a href="stable.gif">
<img src=stable_static.gif></a>
<br>
<a href="stable.gif">[Animated GIF] </a>
<a href="stable.mpg">[MPEG] </a>
<br>
<br>
<STRONG>Figure 7:</STRONG>
\end{rawhtml}
A solution to equations (\ref{K}) and (\ref{C}) 
when the two-dimensional front is stable to transverse perturbations.
The frames in the top row show the time evolution of 
$C(s)$ (red line) and $\kappa(s)$ (blue line). 
The solution converges to a constant
speed flat ($\kappa=0$) traveling front.  The frames in the bottom
row display the corresponding dynamics of the front in 
the physical two-dimensional plane.
\begin{rawhtml}
<BR>
<BR>
<BR>
</center>
\end{rawhtml}
\end{htmlonly}
Crossing the transverse instability boundary causes perturbations
on the front to grow.  
Near the NIB bifurcation, the growing
curvature triggers the ``nucleation'' of a front segment with opposite
velocity as shown in \hyperref{Figure 8}{Fig. }{}{figure:ck_nucleate}. 
The front structures in the $C$-$s$ plane, separating segments
with positive and negative velocities, pertain to spiral waves in the physical
two-dimensional plane.
\begin{latexonly}
\begin{figure}
\centering\includegraphics[width=5.5in]{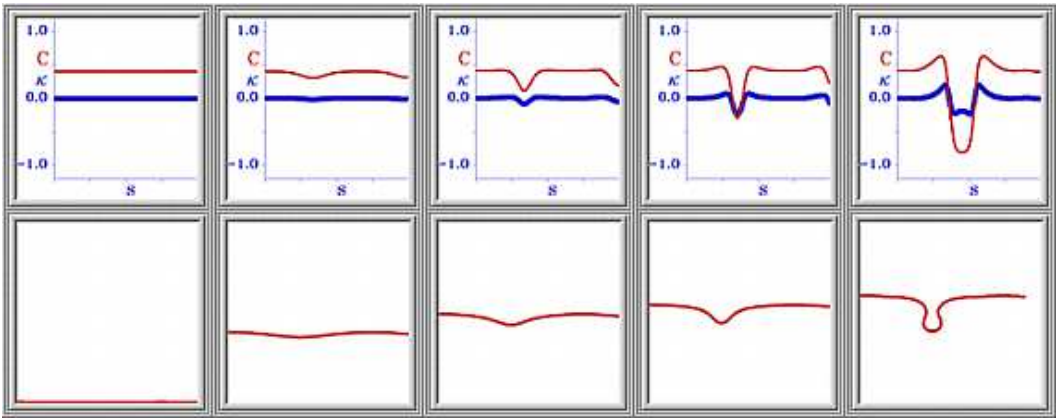}
\caption{
A solution to equations (\protect\ref{K}) and (\protect\ref{C}) 
when the two-dimensional front is unstable to transverse perturbations.
The frames in the top row show the evolution of  $C(s)$ (thin line) and
$\kappa(s)$ (thick line).
The frames in the bottom
row display the corresponding dynamics of the front line in the
physical two-dimensional plane.
A small curvature perturbation grows and the negative curvature
triggers the formation of a region where the front speed
becomes negative.  The boundary points of this region form 
the cores of new rotating spiral waves.
}
\label{figure:ck_nucleate}
\end{figure}
\end{latexonly}

\begin{htmlonly}
\label{figure:ck_nucleate}
\begin{rawhtml}
<br><br>
<center>
<a href="nucleate.gif">
<img src=nucleate_static.gif></a>
<br>
<a href="nucleate.gif">[Animated GIF] </a>
<a href="nucleate.mpg">[MPEG] </a>
<br>
<br>
<STRONG>Figure 8:</STRONG>
\end{rawhtml}
A solution to equations (\ref{K}) and (\ref{C}) 
when the two-dimensional front is unstable to transverse perturbations.
The frames in the top row show the evolution of  $C(s)$ (red line) and
$\kappa(s)$ (blue line).
The frames in the bottom
row display the corresponding dynamics of the front line in the
physical two-dimensional plane.
A small curvature perturbation grows and the negative curvature
triggers the formation of a region where the front speed
becomes negative.  The boundary points of this region form 
the cores of new rotating spiral waves.
\begin{rawhtml}
<br><br>
</center>
\end{rawhtml}
\end{htmlonly}

\section{Conclusion}

We have presented a new mechanism for spontaneous spiral-wave nucleation in
bistable media that leads to spatio-temporal disorder. Unlike most other 
mechanisms which involve destabilization of periodic patterns (see however
\cite{BFHNEE:94,MPSS:96}), this mechanism involves destabilization of 
{\em fronts} and may induce spatio-temporal disorder from a single front 
state.  The dynamical equations, (\ref{K}) and (\ref{C}),
for the front speed and curvature, 
describe the asymptotic behavior of fronts near the front
bifurcation and the transverse instabilities. These equations capture the
process of spiral-wave nucleation and can be used to analyze the transition
from the stable curvature dynamics shown in 
\hyperref{Figure 7}{Fig. }{}{figure:ck_stable}
to dynamics involving nucleation events shown in 
\hyperref{Figure 8}{Fig. }{}{figure:ck_nucleate}.
It would be interesting to find if an
intermediate parameter range exists where the curvature fluctuates 
but no nucleation events occur. 
\section{Appendix}
\label{appendix}

The model used to demonstrate the nucleation of spiral-vortices
in \hyperref{Figure 1}{Fig. }{}{figure:front_turb} is
a doubly-diffusive version of the FitzHugh-Nagumo Equations

\begin{eqnarray}
u_t &=& \epsilon^{-1}(u-u^3-v)+\delta^{-1} u_{xx}\,, \nonumber \\
v_t &=& u-a_1 v -a_0 + v_{xx} \,. \nonumber
\end{eqnarray}
The parameters $a_0$ and $a_1$ are chosen such
that the equations have two stable uniform solutions (bistable)
and $\epsilon$ and $\delta$ control the type and stability
of front solutions between those two stable states.

\hyperref{Figure 9}{Fig. }{}{figure:front_nucleate}
shows a closeup of the middle frame in 
\hyperref{Figure 1}{Fig. }{}{figure:front_turb} 
The transverse instability of the original front solution has
already caused the formation of spiral-vortex pairs
along the front.  The spiral-vortices are identified
with the crossing points of the zero contour
lines of the $u$ and $v$ fields.  At these points
the normal front velocity is zero.
On either side of the spiral-vortex the front propagates in opposite 
directions.

\begin{rawhtml}
<center>
\end{rawhtml}
\begin{figure}[htb]
\htmlimage{align=center,thumbnail=1.4}
\centering\includegraphics[width=4.5in]{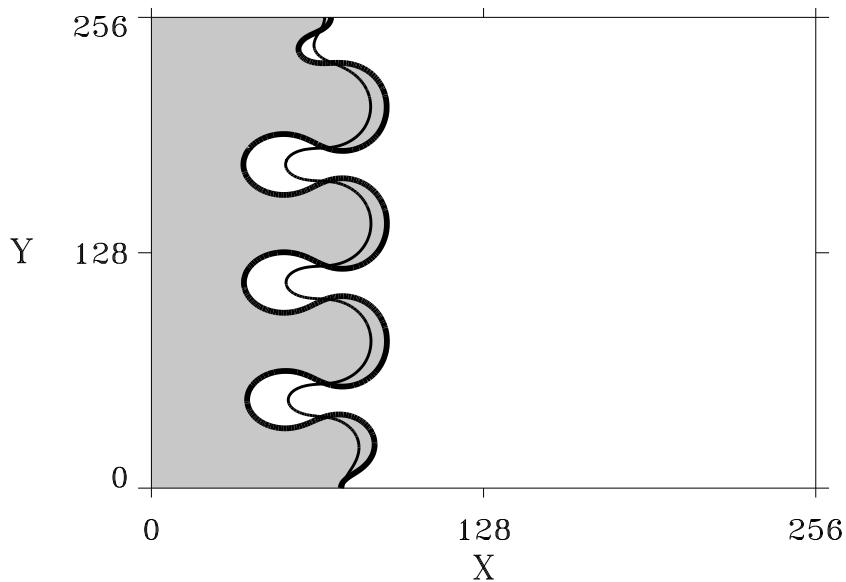}
\caption{
Nucleation of spiral-vortex pairs in the FHN model. 
Each crossing of the zero contour lines
of the $u$ field (thick line) and $v$ field (thin line)
represents a spiral-vortex that forms the core of
a rotating spiral wave.  
}
\label{figure:front_nucleate}
\end{figure}
\begin{rawhtml}
</center>
\end{rawhtml}

\bibliography{reaction}

\end{document}